\documentclass[11pt]{article}

\pdfoutput=1

\usepackage[textwidth=15.5cm,textheight=22.5cm]{geometry}
\usepackage{bm}
\usepackage{amsmath,amssymb,color,graphicx,amscd,amsfonts}
\usepackage{latexsym}
\usepackage{graphicx}
\usepackage{cite}
\usepackage{subfig}
\usepackage{bbm}
\usepackage{ulem}

\usepackage{datetime}

\begin{document}

\thispagestyle{empty}

\setcounter{page}{0}

\begin{flushright} 
SU-ITP-16/06\\
YITP-16-22 \\
LLNL-JRNL-685083

%\today\\[0.2cm]
\yyyymmdddate\today, \currenttime\\[0.2cm]

\end{flushright} 

\vspace{0.1cm}

\begin{center}
{\LARGE

%A microscopic derivation of the thermal Hawking spectrum

A microscopic description of black hole evaporation 

via holography

\rule{0pt}{20pt}  }
\end{center}

\vspace*{0.2cm}

\renewcommand{\thefootnote}{\fnsymbol{footnote}} % Evan changed this because having the first footnote in the text be footnote d was weird.

\begin{center}
	Evan B{\sc erkowitz},$^a$\footnote{
	E-mail: berkowitz2@llnl.gov}
Masanori H{\sc anada},$^{bcd}$\footnote
{
E-mail: hanada@yukawa.kyoto-u.ac.jp} and 
          Jonathan M{\sc altz}$^{eb}$\footnote
          {E-mail: jdmaltz@berkeley.edu or jdmaltz@alumni.stanford.edu}

\vspace{0.3cm}
$^a$ {\it Nuclear and Chemical Sciences Division, Lawrence Livermore National Laboratory, Livermore CA 94550, USA}

$^b${\it Stanford Institute for Theoretical Physics,
Stanford University, Stanford, CA 94305, USA}

$^c${\it Yukawa Institute for Theoretical Physics, Kyoto University,\\
Kitashirakawa Oiwakecho, Sakyo-ku, Kyoto 606-8502, Japan}	

$^d${\it The Hakubi Center for Advanced Research, Kyoto University,\\
Yoshida Ushinomiyacho, Sakyo-ku, Kyoto 606-8501, Japan}

 $^e$
{\it Berkeley Center for Theoretical Physics, University of California at Berkeley,\\
 Berkeley, CA 94720, USA}

\end{center}

\vspace{1cm}

\begin{abstract}
We propose a description of how a large, cold black hole (black zero-brane) in type IIA superstring theory evaporates into 
freely propagating D0-branes, 
by solving the dual gauge theory quantitatively. 
The energy spectrum of emitted D0-branes is parametrically close to thermal when the black hole is large. 
The black hole, while initially cold, gradually becomes an extremely hot and stringy object as it evaporates. As it emits D0-branes, its emission rate speeds up and it evaporates completely without leaving any remnant. 
Hence this system provides us with a concrete holographic description of black hole evaporation without information loss.

\end{abstract}

\begin{center}
Essay written for the Gravity Research Foundation 2016 Awards for Essays on Gravitation.
\end{center}
%%%%%%%%%%%%%%%%%%%%%%%%%%%%%%%%%%%%%%
%%%%%%%%%%%%%%%%%%%%%%%%%%%%%%%%%%%%%%
%%%%%%%%%%%%%%%%%%%%%%%%%%%%%%%%%%%%%%

\newpage

Whether evaporating black holes can be described by a unitary theory via the holographic principle \cite{'tHooft:1993gx,Susskind:1994vu} 
is the key question whose answer should lead to the resolution of Hawking's information paradox \cite{Hawking:1976ra}.
String theory as a proposed theory of quantum gravity has been successful in giving an explanation for the microscopic origin of the Bekenstein-Hawking entropy 
of extremal black holes\cite{Strominger:1996sh}. 
Most believe that a manifestly unitary description of black hole evaporation 
can be obtained by relating superstring theory to a non-gravitational theory via the holographic principle, 
or more concretely via gauge/gravity duality \cite{Maldacena:1997re}. 
However in the almost 20 years since the gauge/gravity duality conjecture, no explicit example of unitary evaporation has been obtained. 
In this essay we propose an explicit scenario, based on concrete calculations in the non-supersymmetric parameter region of gauge theory.

The gauge theory we consider is that of D0-brane matrix quantum mechanics, whose Lagrangian is given by 
\begin{eqnarray}
L
=
\frac{1}{2g_{YM}^2}{\rm Tr}\Bigg\{
(D_t X_M)^2 
+
[X_M,X_{M'}]^2 
+
i\bar{\psi}^\alpha D_t\psi^\beta
+
\bar{\psi}^\alpha\gamma^M_{\alpha\beta}[X_M,\psi^\beta] 
\Bigg\},  
\end{eqnarray}
where $X_M$ $(M=1,2,\cdots,9)$ and $\psi_\alpha$ $(\alpha=1,2,\cdots,16)$ are $N\times N$ bosonic and fermionic 
Hermitian matrices, $D_t$ is the covariant derivative given by 
$D_t=\partial_t -i[A_t,\ \cdot\ ]$ and $A_t$ is the $U(N)$ gauge field. 
$\gamma^M_{\alpha\beta}$ are the left-handed part of the gamma matrices in (9+1)-dimensions. 

\begin{figure}[htbp]
   \begin{center}\rotatebox{0}{
   \scalebox{0.5}{
     \includegraphics[height=6.5cm]{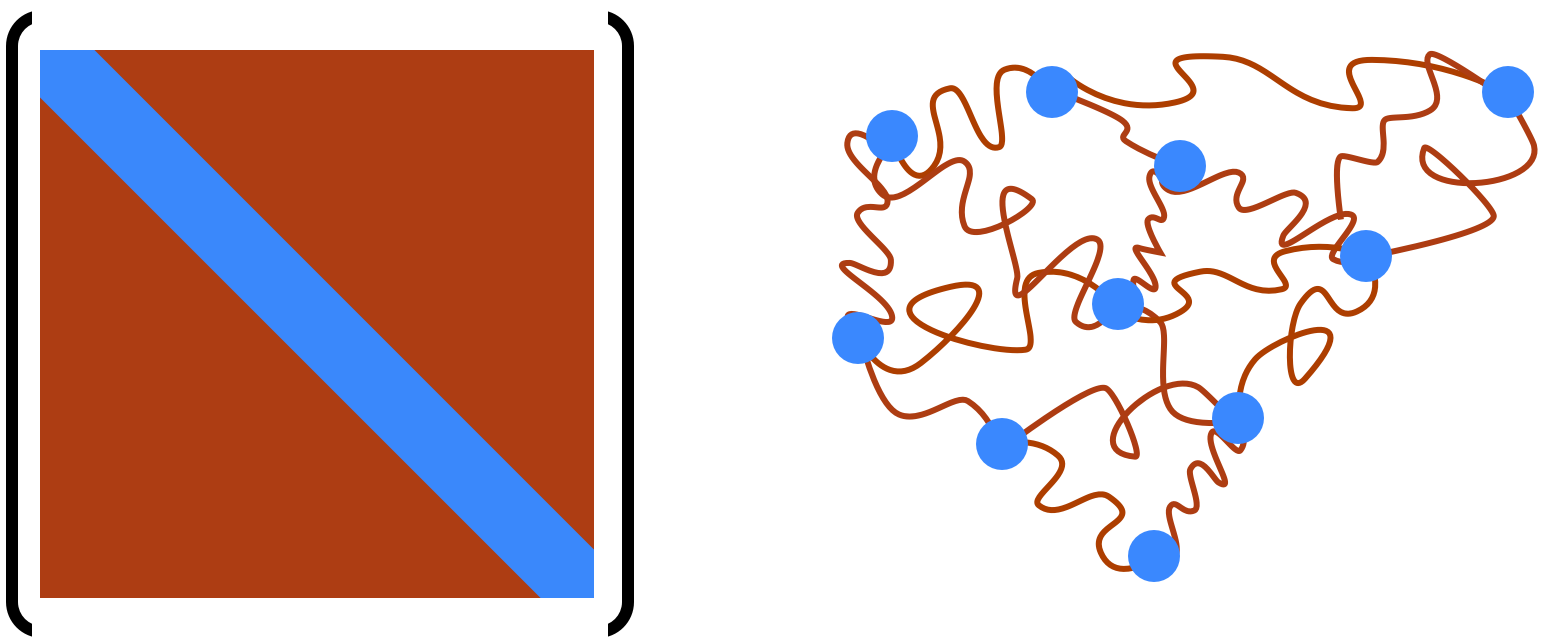}}}
   \end{center}
   \caption{Schematic interpretation of matrix components as D0-branes (diagonal elements) and open strings (off-diagonal elements)\cite{Witten:1995im}.}
\label{fig:interpretation_matrices_1}
\end{figure}

Among several physical interpretations of this model\cite{Banks:1996vh,deWit:1988ig,Witten:1995im,Itzhaki:1998dd}, we consider the one based on the gauge/gravity duality\cite{Itzhaki:1998dd}, which claims that near the 't Hooft large-$N$ limit this model describes type IIA superstring theory. Intuitively, diagonal and off-diagonal elements of matrices are interpreted 
as D0-branes and open strings, respectively (Fig.~\ref{fig:interpretation_matrices_1}). 
A black hole is described by states where all the D0-branes and strings form a single bound state, described by generic non-commuting matrices with all eigenvalues clumped in the neighborhood of a point in space. 
The deviation from strong coupling and $1/N$ corrections are conjectured to describe stringy corrections to the black hole. 
Previous thermodynamic simulations have provided us with quantitative evidence that the duality is valid including stringy corrections \cite{Anagnostopoulos:2007fw,Catterall:2008yz,Kadoh:2015mka,Filev:2015hia,Kabat:1999hp,Kabat:2000zv,Hanada:2008ez,Hanada:2013}.
In fact this is the only case where the duality has been tested at the fully stringy level. This fact motivates us to study this model further, 
in order to obtain an explicit counter-example to information loss.  

The black zero brane seems to be rather different from the Schwarzschild black hole. Firstly,   
the emission of neutral massless particles is suppressed \cite{Hanada:2013} 
because of the near-extremal limit taken in deriving the duality \cite{Itzhaki:1998dd} and secondly it possesses Ramond-Ramond charge which results in it having a positive specific heat
and a stable supersymmetric zero-temperature limit. 
Can we find any resemblance to the usual black hole evaporation that we know of on the gravity side?
As we will see, we can if we take into account the emission of the charges.
In particular, that effect makes the specific heat negative.

\begin{figure}[htbp]
   \begin{center}
     \includegraphics[width=\textwidth]{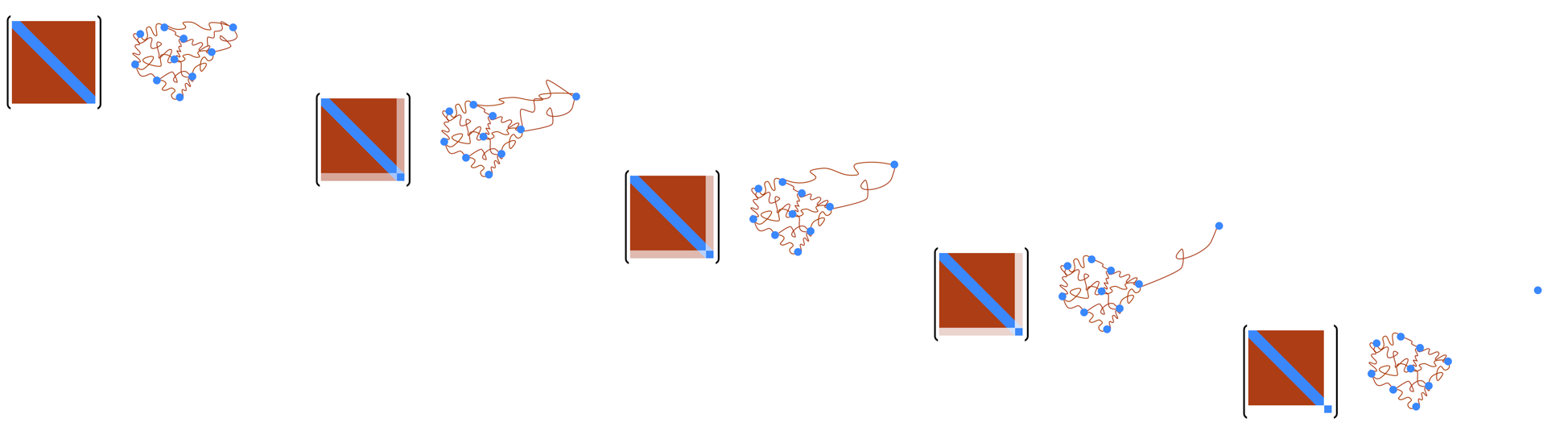}
   \end{center}
   \caption{A cartoon of a single D0-brane escaping the black hole.}
\label{fig:interpretation_matrices_2}
\end{figure}

The key feature of the D0-brane matrix model is the flat directions, i.e. $[X_M,X_{M'}]= 0$, 
which exist both classically and quantum mechanically \cite{de-Wit:1988ct}. 
The configurations corresponding to a black hole have an instability due to these flat directions at finite $N$ \cite{Anagnostopoulos:2007fw},  
which can be understood as the Hawking radiation of D0-branes, which carry charges.
Before emission, matrices are fully noncommutative. When one D0-brane is emitted, 
they take block diagonal forms, 
\begin{eqnarray}
X^M
=
\left(
\begin{array}{cc}
X_{\rm BH}^M & 0 \\
0 & x_{\rm D0}^M
\end{array}
\right),  
\label{block_diagonal}
\end{eqnarray}
where $X_{\rm BH}$ are $(N-1)\times (N-1)$ (see Fig.~\ref{fig:interpretation_matrices_2}). 
Because \eqref{block_diagonal} is of this special form, the emission is entropically suppressed from the generic case. 
However, because of ergodicity, such rare configurations will eventually be realized. 
Once a D0-brane is emitted -- namely, \eqref{block_diagonal} is realized and the D0-brane goes sufficiently far from the black hole -- the flat direction opens up and 
the D0-brane can go infinitely far along it.
We have shown that the black hole becomes hotter as it evaporates via this emission process \cite{Berkowitz:2016znt}. 
Before the particle emission, there were $N^2$ degrees of freedom. 
However, after the emission, the number of degrees of freedom decreases to $(N-1)^2+1$, because the off-diagonal blocks decouple.
Due to energy conservation, the amount of energy per degree of freedom ($\simeq$ temperature) goes up.\footnote{
Here we implicitly assumed that the emitted particle carries only $O(N^0T)$ energy. This assumption will be justified in \eqref{eq:spec} by showing the spectrum of radiation follows the Boltzmann distribution. 
The change of the temperature will also be calculated explicitly. }
This mechanism can hold for any black hole described by matrices; it is a generic feature of string theory and gauge theory. 
Also note that this is a physically meaningful process, 
because the time scale for the emission is $\sim e^N$, as we can see by comparing the entropies, 
which is smaller than the recurrence time of the black hole $\sim e^{N^2}$.

Based on this observation, we would like to understand the entire evaporation process in this essay. 
Firstly, let us determine the energy spectrum of the radiation.
Let us consider a matrix configuration describing a black hole and one particle of radiation. 
When the radiation goes parametrically far from the black hole, it has the block-diagonal form explained above. 
By assuming the mechanism of evaporation proposed in \cite{Berkowitz:2016znt}, which relies only on generic properties of a chaotic system, 
it is reasonable to assume that all possible final states appear with equal weights. 
Then, we only have to count the number of states as a function of $E_{\rm rad}$, by fixing the total energy of the system  
$E=E_{\rm BH}+E_{\rm rad}$. 
As long as $N\gg 1$, the derivation of the spectrum is essentially the same as 
the derivation of the canonical ensemble from the microcanonical ensemble.
The density of states $W$
\begin{equation}
W=\sum_{E_{\rm BH}+E_{\rm rad}=E}W_{\rm BH}(E_{\rm BH}) \cdot W_{\rm rad}(E_{\rm rad}), 
\end{equation}
where $W_{\rm BH}=e^{S_{\rm BH}}$, 
can be evaluated as 
\begin{equation}
W
=
\sum_{E_{\rm rad}}\left(
W_{\rm rad}(E_{\rm rad}) \cdot e^{S_{\rm BH}(E-E_{\rm rad})}
\right)
\simeq
W_{\rm BH}(E) 
\sum_{E_{\rm rad}}\left(
W_{\rm rad}(E_{\rm rad}) \cdot e^{-\frac{E_{\rm rad}}{T_{\rm BH}(E)}}
\right), \label{eq:spec}
\end{equation}
where $T_{\rm BH}\equiv\left(\frac{dS_{\rm BH}}{dE}\right)^{-1}$. 
Here, corrections of order $1/N$, which contain information about quantum gravity effects, have been ignored. 
In principle we can calculate such corrections by fully solving the matrix model. 

In type-IIA parameter region, an entropy argument shows that D0-branes are emitted one by one 
---the timescale for $k$-particle emission is $\sim e^{kN}$ while $N$ single-particle emissions still only take time $\sim e^{N}$, so that the black hole will entirely evaporate by single particle emissions.
Then, $E_{\rm rad}$ is the kinetic energy of the emitted D0-brane.

Consider a black zero-brane with energy $E_0=O(N^2)$. 
In the high and low temperature regions, the change of the temperature can be determined analytically. 
Here, the energy $E$, temperature $T$, and the charge $N$ are related by $E(N,T)=6N^2T$ (low temperature) 
and $E(N,T)=7.41(g_{YM}^2N)^{-3/5}T^{14/5}$, respectively. (See e.g. \cite{Anagnostopoulos:2007fw} and references therein.)
After the emission of a D0-brane, the temperature 
changes to $T'=T+\Delta T$. From the energy conservation $E(N,T)=E(N-1,T')+E_{\rm rad}$, 
and $E_{\rm rad}\sim O(N^0)$, it can be shown that $\Delta T=2T/N$ (low temperature) and $\Delta T=T/2N$ (high temperature), 
regardless of the value of $E_{\rm rad}$. 
In this manner, the black zero-brane heats up as it emits D0-branes, because the majority of the energy is 
left in the black hole while the number of dynamical degrees of freedom decreases\cite{Berkowitz:2016znt}.
When the number of D0-branes left in a black hole $n$ becomes smaller, 
the weak-coupling description becomes valid (it happens when $n$ is still $N$ times an order one value) 
and beyond that point the temperature scales as $T\sim E_0/n^2$.\footnote{
Compared to $E_0$, the amount of the energy carried away by the radiations is 
$\frac{1}{N^2}+\frac{1}{(N-1)^2}+\cdots\frac{1}{(n+1)^2}\sim \frac{1}{n}$, which is negligible as long as $n$ is large.
}
(When $n\lesssim\sqrt{N}$, the temperature is as high as $N$, which is the order of D0-brane mass. 
We assume the duality is valid even at such high temperature. We note that this is an intricate issue  
related to the order of large-$N$ and near-extremal limits.)
This simple observation has immense impact -- at high temperature, the dynamics of the matrix model 
is weakly coupled and can be fully studied by using standard numerical tools \cite{Asplund:2012tg,Gur-Ari:2015rcq,Berkowitz:2016znt}, 
and hence, the study of the very last moment of 
the black hole evaporation, with full quantum gravity effects, is within reach. 

What about the emission rate? D0-branes can escape from the black hole when the flat direction opens up. 
This distance should be proportional to $T$, beyond which the strings are too heavy to be thermally excited. 
As a rough estimate, at high temperature, let us assume that the classical approximation is valid below the distance $T$, 
and that D0-branes can travel freely beyond there. 
Then the time scale for this to happen can be calculated by using the eigenvalue distribution 
of the classical theory \cite{Gur-Ari:2015rcq,Berkowitz:2016znt}, 
\begin{eqnarray}
\rho(r)
\sim
(g_{YM}^2nT)^{-1/4}
\cdot
\left(
\frac{r}{(g_{YM}^2nT)^{1/4}}
\right)^{-8(2n-3)}, 
\end{eqnarray}
where $r\equiv|\vec{x}_{\rm D0}|$.
The time scale for $r$ to reach $T$ can be estimated from the volume of the phase space as
\begin{eqnarray}
\left(
\int_T^\infty dr\rho(r)
\right)^{-1}
\sim
%\left(\frac{g^2_{YM}E_0}{n}\right)^{1/4}
%\left(\frac{g_{YM}^{-1/2}E_0^{3/4}}{n^{7/4}}
%\right)^{8(2n-3)-1} 
%\sim
n\cdot\left(\frac{N}{n}\right)^{28n-175/4}. 
\end{eqnarray}
Therefore the emission rate increases as the black hole becomes smaller; 
it changes from the exponential of $N$ to a power of $N$. 
The black zero-brane ends up as freely propagating D0-branes. The black zero-brane is just a resonant state; the most entropically favorable state, non-interacting D0-branes, will eventually be realized. 
There is no information loss: the final state knows everything about the initial condition because the time evolution is unitary. 

There still remain challenging problems on which this model provides 
an ideal playground for their resolution: Can a Schwarzschild black hole in M-theory be described? Is it possible to see quantum entanglement? 
How is the geometry of the black hole encoded into the matrices? 
We hope to report progress along such directions in the future publications, by combining numerical experiments and theoretical considerations.

\begin{center}
{\bf Acknowledgments}
\end{center}
The authors would like to thank H.~Shimada, especially for discussion on the thermal nature of the radiation spectrum.
We would also like to thank 
Guy Gur-Ari, Yasunori Nomura, Enrico Rinaldi, Stephen Shenker, Leonard Susskind and Masaki Tezuka for discussions and comments.
The work of M.~H. is supported in part by the Grant-in-Aid of the Japanese Ministry 
of Education, Sciences and Technology, Sports and Culture (MEXT) 
for Scientific Research (No. 25287046). 
This work was performed under the auspices of the U.S. Department of Energy by Lawrence Livermore National Laboratory under Contract DE-AC52-07NA27344.
The work of J.~M. is supported by the California Alliance fellowship (NSF grant 32540).

%%%%%%%%%%%%%%%%%%%%%%%%%%%%%%%%%%%%%%%%%%%%%%

\end{document}